\documentclass[12pt,preprint]{aastex}
\usepackage{color}
\usepackage{natbib}

\slugcomment{Accepted for Publication in ApJ}

\begin{document}
\shorttitle{HD~207832: A New Exoplanetary System}
\shortauthors{Haghighipour et al.}

\title{The Lick-Carnegie Survey: A New Two-Planet System Around the Star HD~207832}
\author{Nader Haghighipour\altaffilmark{1}, R. Paul Butler\altaffilmark{2},
Eugenio J. Rivera\altaffilmark{3}, 
Gregory W. Henry\altaffilmark{4},
and Steven S. Vogt\altaffilmark{3},}

\altaffiltext{1}{Institute for Astronomy and NASA Astrobiology Institute,
University of Hawaii-Manoa, Honolulu, HI 96822}
\altaffiltext{2}{Department of Terrestrial Magnetism, 
Carnegie Institute of Washington, Washington, DC 20015}
\altaffiltext{3}{UCO/Lick Observatory, Department 
of Astronomy and Astrophysics,
 University of California at Santa Cruz, Santa Cruz, CA 95064}
\altaffiltext{4}{Center of Excellence in Information Systems,
Tennessee State University, Nashville, TN 37209}

\begin{abstract}
Keck/HIRES precision radial velocities of HD 207832 indicate the presence of two Jovian-type planetary
companions in Keplerian orbits around this G star. The planets have minimum masses of  
$M \sin i = 0.56 \, {M_{\rm Jup}}$ and $0.73 \, {M_{\rm Jup}}$, with orbital periods of 
$\sim 162$ and $\sim 1156$ days, and eccentricities of 0.13 and 0.27, respectively. 
Str\"omgren $b$ and $y$ 
photometry reveals a clear stellar rotation signature of the host star with a period of 
17.8 days, well separated from the period of the radial velocity variations, 
reinforcing their Keplerian origin. The values of the semimajor axes of the planets suggest that
these objects have migrated from the region of giant planet formation to closer orbits. 
In order to examine the possibility of the existence of additional (small) planets in the system,
we studied the orbital stability of hypothetical terrestrial-sized objects in the region between the 
two planets and interior to the orbit of the inner body. Results indicated that stable 
orbits exist only in a small region interior to planet b. However, the current
observational data offer no evidence for the existence of additional objects in this system. 
\end{abstract}


\section{Introduction}
The planetary census has currently exceeded an impressive 750 extrasolar planets. 
Planetary companions have been successfully detected using a variety of techniques, primarily 
radial velocity and transit photometry, with more than 700 and close to 230 planets 
detected by each method, respectively.
Other successful techniques include microlensing \citep[15 planets, see e.g.,][]{Batista11}, 
astrometry \citep[e.g.,][]{Muterspaugh10}, stellar pulsations \citep{Silvotti07}, 
direct imaging \citep[31 planets, see e.g.,][]{Chauvin05,Kalas08,Marois10}, and the transit timing
variation method \citep[16 planets, see e.g.,][]{Holman10,Lissauer11,Doyle11,Welsh12}.

The radial velocity method has been used to characterize $\sim 92\%$ of all known planets,
and continues to be the dominant technique. Both its continued
productivity and its ability to accurately probe
planetary architectures into the vicinity of the terrestrial-mass region
\citep[e.g.,][]{Rivera05,Mayor09,Vogt10,Anglada12} are a testament to the capability of this technique
and its rapid technological advances.
For the past 18 years, we have used this technique and monitored a large number of nearby stars 
with the High Resolution Echelle Spectrometer (HIRES) at the Keck observatory. In this paper, 
we present new radial velocity (RV) and photometric observations for one of our 
target stars: HD~207832. 

The plan of this paper is as follows. In \S 2, we discuss the basic properties of HD~207832.
In \S 3, we describe the new radial velocities, derive a Keplerian two-planet model of the system, 
and describe the new APT observations. Finally, in \S 4, we discuss the properties of this
new planetary system and the possibility of its hosting other planetary bodies.

\section{HD 207832 Properties}
 
HD~207832 (also known as CD-2615858, CPD-267292, SAO190699, 2MASSJ21523626-2601352, 
HIP 107985, and TYC6956-00378-1) is a G5 dwarf with a visual magnitude of 8.786$\pm$0.014.
We present in Table~1 a few basic parameters of this star.
Unless otherwise noted, the data are as listed in the SPOCS
\citep{Valenti05} and the NASA NStED databases\footnote{http://nsted.ipac.caltech.edu/}.

HD 207832 has a  metallicity of [Fe/H]=0.06 and an age $<$ 4.5 Gyr \citep{Holmberg09}.
The parallax of this star, as revised by \citet{VanLeeuwen07}, is 18.37$\pm$0.92 mas, 
which corresponds to a distance of 54.4$\pm$2.7 pc. As a result, HD~207832 has an absolute visual 
magnitude ($M_{\rm v}$) of 5.11$\pm$0.11. The bolometric correction of HD~207832 in the visual 
is -0.080$\pm$0.055 \citep{Masana06} implying an absolute bolometric magnitude of 5.03$\pm$0.12. 
Assuming 4.75 for the corresponding quantity for the Sun, this bolometric magnitude points to 
a luminosity of 0.773$\pm$0.085 $L_{\odot}$ for this G star. 

The simple mass-luminosity relationship for main sequence stars, $M=L^{1/3.9}$,
indicates that the mass of HD~207832 is approximately 0.94$\pm$0.10 $M_{\odot}$. This is 
in good agreement with the value determined from interpolation tables in \citet{Gray92}
and the value of 0.97 $M_{\odot}$ as reported by \citet{Nordstroem04}.
HD 207832 has a semi-diameter of 0.077$\pm$0.001 mas (Masana et al. 2006), which given 
its distance of 54.4 pc, indicates a stellar radius of 0.901$\pm$0.056 $R_{\odot}$.
Using the above-mentioned values of mass and radius, we find a simplistic estimate for
$\log{g}=4.502 \pm 0.071$, in good agreement with the expected
range of values for a main sequence G star. The effective temperature of
HD 207832, obtained from the Stefan-Boltzmann law with the luminosity and radius as
mentioned above, is 5710$\pm$81 K, which is in rough agreement with 5649 K as reported
by Holmberg et al. (2009).

The value of $V\sin{i}$ for HD~207832 is approximately 3 km\,s$^{-1}$  \citep{Nordstroem04}. 
From our photometry presented in the next section, we determined a rotation period of 17.8 days.
While obtaining the RV's presented here, we measured the Mt. Wilson $S$ index
and found that it has a mean value of 0.258. We also measured $\log{{\rm R'}_{\rm HK}}=-4.62$. 
These values are similar to $S= 0.207$ and $\log{{\rm R'}_{\rm HK}}=-4.80$, as in 
\citet{Jenkins08}.

\section{New Radial Velocity and Photometric Observations}

\subsection{Radial Velocities}

The HIRES spectrometer \citep{Vogt94} of the Keck-I telescope was used for all the new RV's 
presented in this paper. Doppler shifts were measured in the usual manner \citep{Butler06} 
by placing a gaseous Iodine absorption cell just ahead of the spectrometer slit in the converging 
beam from the telescope. This Iodine cell superimposes a rich forest of 
Iodine lines on the stellar spectrum, providing a wavelength calibration and proxy for the 
point spread function (PSF) of the spectrometer. The Iodine cell is sealed and 
temperature-controlled to 50 $\pm$ 0.1 C such that the column density of Iodine remains 
constant.  For the Keck planet search program, we operate the HIRES spectrometer at a spectral 
resolving power R $\approx$ 70,000 and wavelength range of 3700-8000\,\AA, though only the 
region 5000-6200\,\AA\ (with Iodine lines) is used in the present Doppler analysis. The Iodine 
region is divided into $\sim$700 wavelength intervals of 2\,\AA\ each. Each interval produces 
an independent 
measure of the wavelength, PSF, and Doppler shift. The final measured velocity is the weighted 
mean of the velocities of the individual intervals. All radial velocities have been corrected to 
the solar system barycenter, but are not tied to any absolute radial velocity system. As such, 
they are ``relative'' radial velocities.

Table~2 lists the complete set of 86 relative RV's for HD~207832, corrected to the 
solar system barycenter. We present results using only the internal uncertainties.
The median internal uncertainty for our observations is 1.95 m\,s$^{-1}$, the peak-to-peak 
velocity variation is 95.84 m\,s$^{-1}$, and the velocity scatter around the mean RV 
is 21.29 m\,s$^{-1}$.
The internal uncertainties quoted for all the RV's in this paper reflect only one term in the 
overall error budget and result from a host of systematic errors such as characterizing and 
determining the PSF, detector imperfections, optical aberrations, effects of under-sampling 
the Iodine lines, etc. Two additional major sources of error are photon statistics and stellar jitter.
The latter, which varies substantially from star to star, can be mitigated to some degree by 
selecting magnetically-inactive older stars and by time-averaging over the star's unresolved 
low-degree surface p-modes. For HD 207832, the expected jitter is 4.22 m\,s$^{-1}$ \citep{Isaacson10}.  
This is in accord with the modest level of activity suggested by the quoted values of $\log{{\rm R'}_{\rm HK}}$.
All observations have been further binned on 2-hour timescales.

Figure 1 shows the results.
The top panel of this figure shows the individual RV observations, and the 
middle panel shows the weighted Lomb-Scargle (LS) periodogram of the full RV data set
\citep{Gilliland87}. In generating this periodogram, we used only the internal uncertainties
in the statistical weights and did not include jitter. To examine the effect of jitter, we added in
quadrature 4.22 m\,s$^{-1}$ to each internal uncertainty and reproduced the periodogram. The  
results showed only negligible differences.

\begin{figure}
\center
\includegraphics [scale=0.31, angle=270]{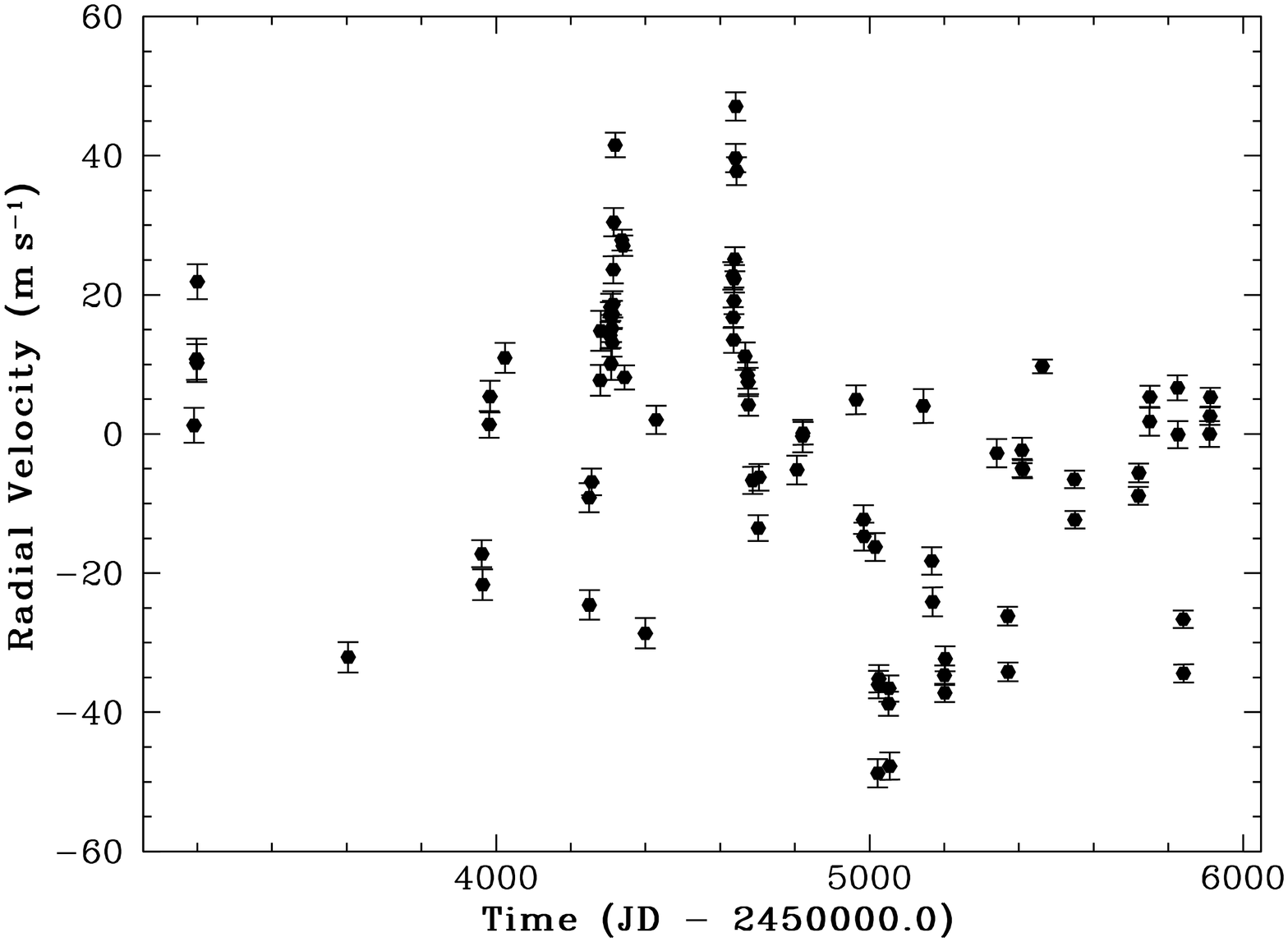}
\includegraphics [scale=0.31, angle=270]{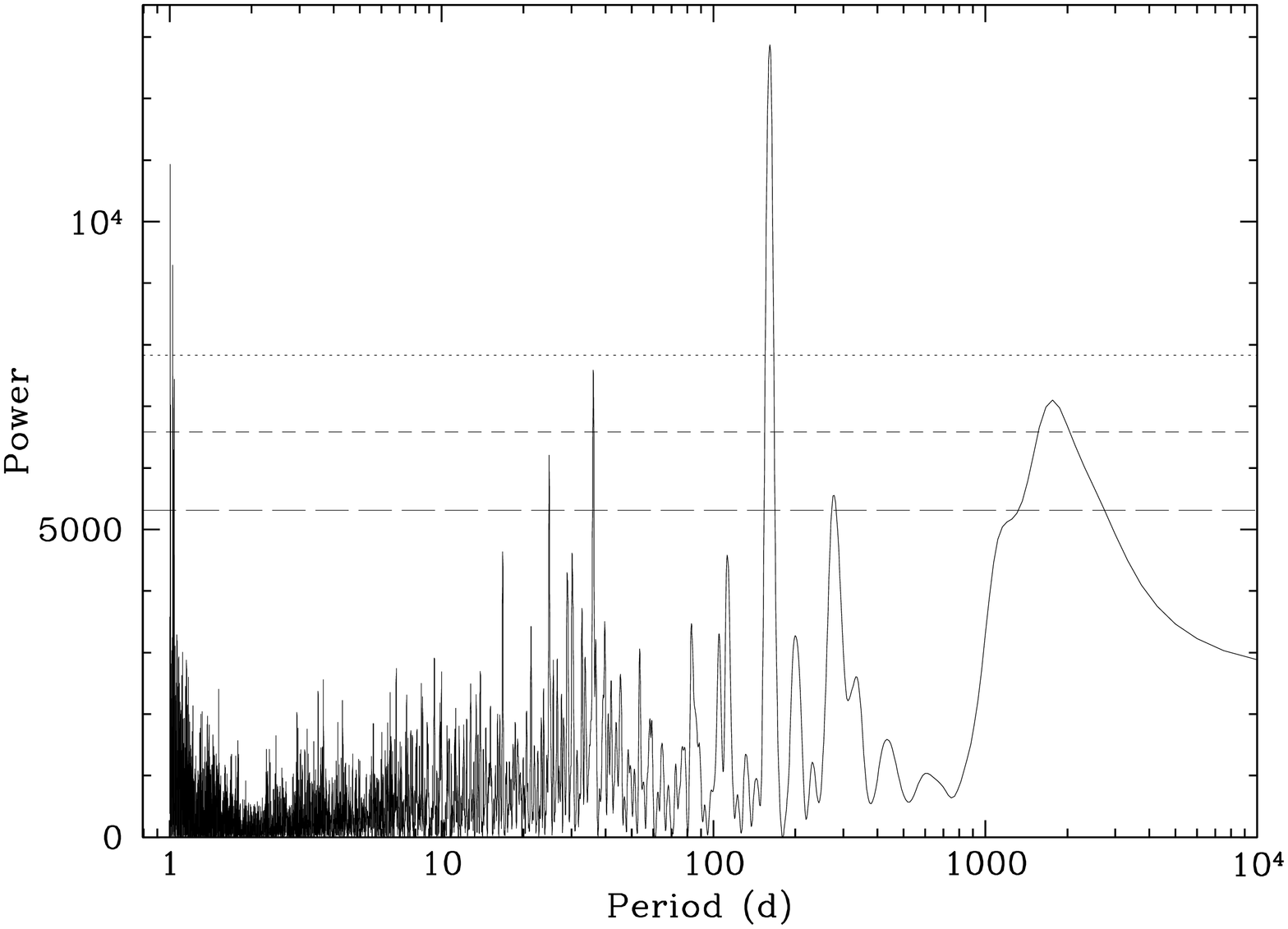}
\includegraphics [scale=0.31, angle=270]{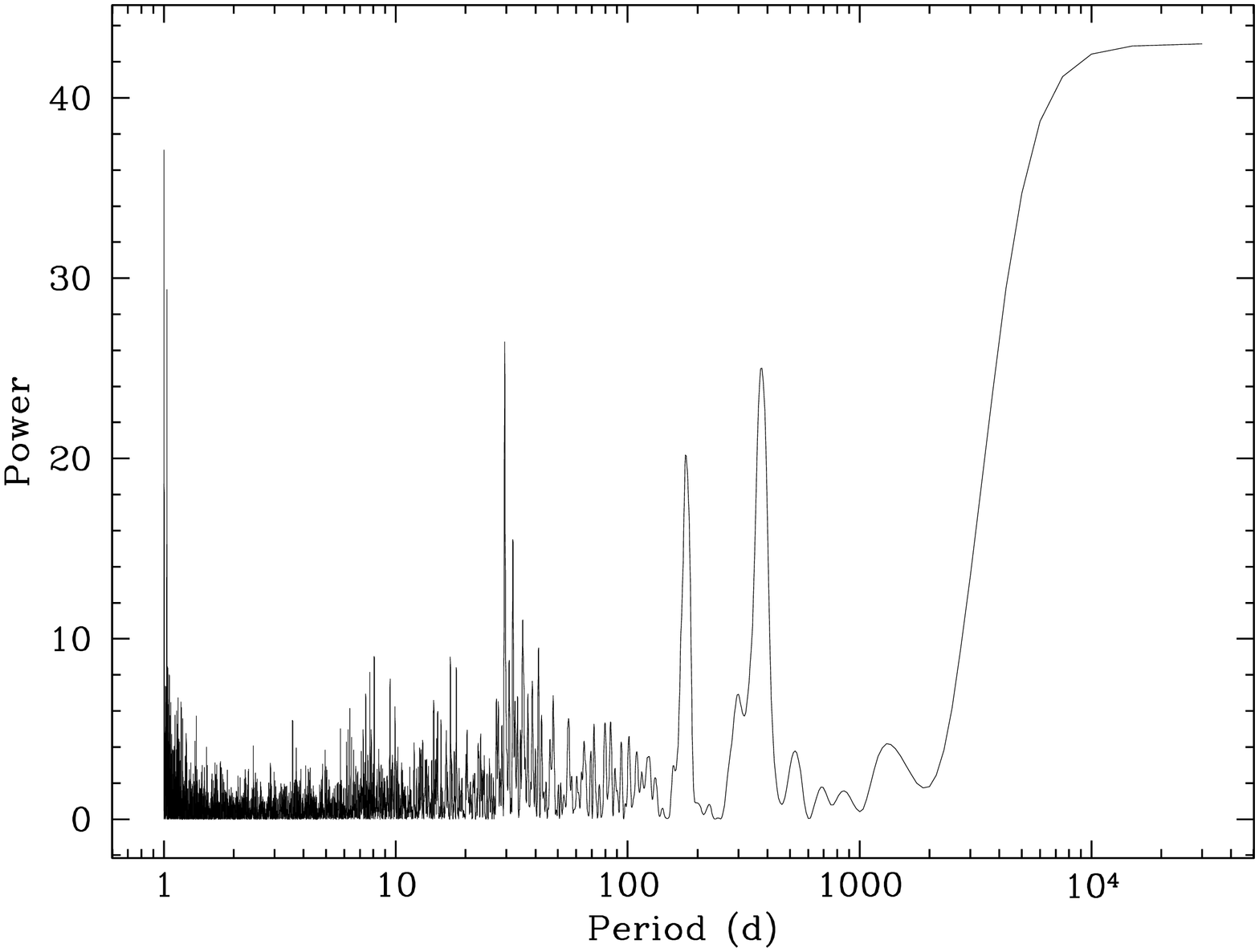}
\caption{Radial velocity data and periodograms for HD~207832. \textit{Top panel:} Relative 
HIRES/Keck radial velocity data. \textit{Middle panel:} Weighted Lomb-Scargle periodogram of
 the radial velocity data. \textit{Bottom panel:} Power spectral window.}\label{fig:data_HD207832}
\end{figure}

The three horizontal lines in this figure and other comparable plots represent, from top to bottom, 
the 0.1\%, 1.0\%, and 10.0\% False Alarm Probability (FAP) levels, respectively. The 
algorithm for computing the LS periodogram is described in detail in \S 13.8 of Press et al. (1992).
To estimate our FAP levels, we randomly assign (without replacement) an observed RV -- or residual
RV when we consider the residuals for a fit -- (with its corresponding uncertainty) to each
observing epoch.  We repeat this $10^5$ times.  For each synthetic RV (or residual RV) set generated
in this manner, we compute the weighted LS periodogram. Our FAP estimates are then the fraction
of these periodograms in which the power in the tallest peak equals or exceeds the power in the
periodogram of the real RVs (or residual RVs). We do this to circumvent potential problems associated
with the application of the FAP estimation method(s) in Press et al. (1992), which are
strictly applicable only in the case of single, isolated signals in the presence of Gaussian noise
with known variance \citep{Koen90}.
For the strong Keplerian signal at $P\sim$162 days in the RV data set, we find a FAP $<10^{-5}$.
Finally, the lower panel of Figure~\ref{fig:data_HD207832} shows the power spectral window.
This spectral window indicates spurious power that might  be introduced into the data  
from the choice of sampling times alone, and it can be used to aid in the identification of aliased signals.

\subsection{Keplerian Solution}

Table 3 summarizes a one-planet Keplerian fit for HD 207832. Orbital fits were derived 
using the Systemic Console \citep{Meschiari09}\footnote{Downloadable at \url{http://www.oklo.org}}.
The errors on each parameter are estimated using the bootstrap technique with 1000  
realizations of the RV data sets. We fit all the realizations, and the uncertainties in the
best-fit parameters are determined by the ranges in each parameter containing 68.2\% of the distributions of the fitted values.
For each planet, we list best-fit period 
($P$), eccentricity ($e$), semi-amplitude ($K$), 
longitude of pericenter ($\varpi$), mean-anomaly (MA), minimum mass ($M \sin{i}$) 
and semi-major axis ($a$). 

The dominant peak in the periodogram of the RV's is well fitted with a Keplerian orbit of 
period 161.82
days and semi-amplitude $K = 21.30$ m\,s$^{-1}$ (top panel of Figure~2). 
Together with the assumed stellar mass of 0.94 $M_{\odot}$,
this amplitude suggests a planet with a minimum mass  
of $M\sin{i} = 0.62 \, M_{\rm {Jup}}$. The best-fit orbit for this planet is moderately eccentric 
($e \approx 0.18$). 
This one-planet fit achieves $\chi_{\nu}^2 = 43.99$, with an RMS of 12.33 m\,s$^{-1}$.
If we add in the estimated jitter and perform a new fit, the only significant difference 
will be the reduction of $\chi_{\nu}^2$ to 8.22. Within the uncertainties, the two fits were indistinguishable.

The bottom panel of Figure2 shows the periodogram of the residuals 
to the single-planet fit. %and the corresponding FAPs. 
The dominant peak at $P = 1111.2$ days with a FAP 
$2.3 \times 10^{-4}$ is indicative of a secure detection of an additional planet.
Again, we find that the addition of the expected jitter to the internal uncertainties
has a minor effect on this periodogram.
The peak near 30 days
is an alias of the 1111-day peak. In fact, the 30, 150, and 1111-day peaks in the one-planet
residuals periodogram are all inter-related via the lunar synodic and 1/2-year peak in
the Power Spectral Window of Figure~1 (bottom panel).

Our best combined two-planet fit indicates a second planet with $P = 1155.70$ days, 
$K = 15.34$ m\,s$^{-1}$, and a minimum mass of 
$M \sin {i} = 0.73 \, M_{\rm {Jup}}$ (Table~4). The best-fit orbital model of the second planet suggests a
moderate eccentricity, ($e \approx 0.27$). With this revised fit, 
we obtain $\chi_{\nu}^2 = 22.93$ and an RMS of the residuals of 8.43 m\,s$^{-1}$.
The F-test of the two-planet fit vs. the one-planet fit gives a probability 
of $5 \times {10^{-4}}$ (based on the difference in RMS values\footnote{This value reduces drastically to 
$3 \times {10^{-10}}$ when calculated based on the difference in $\chi_{\nu}^2$.})
that the two-planet fit is not significantly different from
the one-planet fit, indicating further evidence in support of the two planet model.
Similar to the case of the one-planet fit, the only significant result of adding in the expected jitter 
is that for this two-planet model, $\chi_{\nu}^2$ will reduced to 4.05. As with the one-planet model, within the uncertainties, the 
fitted parameters for fits with and without jitter have negligible differences.

Figure 3 shows phase folded RV's for the two-planet fit. 
The top panel corresponds to the case where the period of the inner planet has been used 
and the effect of the outer planet has been subtracted. Similarly, the middle panel 
is for the case where the period of the outer planet was used and the effect of the
inner planet was subtracted.
The bottom panel of this figure shows the periodogram of the residuals of the best-fit solution. 
The FAP of the tallest peak is $\sim50\%$. The corresponding result when we added in the
expected jitter was similar, although the period of the tallest peak was slightly different.
The current data set thus offers no compelling evidence for additional planets.\footnote{As 
a check of the reliability of our results, we calculated FAPs for our one- and two-planet
fits as well as for the most prominent signal in the residuals, using the formalism presented by Baluev (2008).
In agreement  with our previous calculations, we found that for the one-planet fit, FAP $<10^{-16}$, 
for the two-planet fit, it is $<10^{-5}$, and for the residuals, it is $<0.5$.
The periodograms for each case were equivalent to those shown in Figures 1-3, confirming the secure
detection of the two planets.}

The periodograms of the residuals of the two-planet fits
(with and without the addition of the expected jitter) contain peaks
(of little significance) with periods ranging from 18 to 22 days.
Also, the analysis of the Mt. Wilson $S$ index indicates a periodicity of
$\sim$ 19 days.  These values are in rough agreement with the photometric period
discussed in the next section.  Inspection of the individual seasons of the $S,$ RV,
and photometric observations indicates that there is significant variation in the
peak period (and its associated power) in the periodograms of these quantities.
These weak periodicities may be tied to the rotation period of the star
through the presence of stellar spots and chromospheric activity.  The lack of a strong
coherent signal over the full time spans of the observations and the relatively
large power levels in the periodogram of the two-planet residuals suggest the
presence of additional (non-Gaussian) noise.  Since the FAP for the most prominent
signals in the residuals of the two-planet fit are $\sim50\%$ and there is evidence
for additional noise which does not show strong coherence over the time span of
the observations, we do not attempt to fit for the stellar rotation (and any other
potential remaining signals).

\subsection{APT Photometry}

In addition to the Keck radial velocities, we obtained Str\"omgren $b$ and 
$y$ photometric observations with the T12 0.80~m automatic photometric 
telescope (APT) at Fairborn Observatory in Arizona.  The T12 APT uses a 
two-channel precision photometer with two EMI 9124QB bi-alkali photomultiplier 
tubes to make simultaneous measurements in the two passbands. We programmed 
the APT to make differential brightness measurements of HD~207832 
($V=8.78$, $B-V=0.69$, G5 V) with respect to three comparison stars:  
HD~207760 ($V=6.19$, $B-V=0.37$, F0 V), HD~206797 ($V=7.35$, $B-V=0.40$, 
F2 III), and HD~208483  ($V=7.64$, $B-V=0.48$, F4 V). A detailed description 
of the automatic telescope, precision photometer, observing procedures,
data reduction, calibration, and photometric precision can be found in
\citet{Henry99}.  The typical precision of a single observation is approximately
0.0015 mag on good nights.

The T12 APT acquired 310 differential observations of HD~207832 during the
2007 -- 2010 observing seasons.  The observing seasons are short, two or
three months, because the star's declination is $-26\arcdeg$, making it 
difficult to observe from Arizona.  Also, the star comes to opposition 
in mid-August when the Fairborn APT site is forced to shut down for two 
months by the annual monsoons.  To improve the photometric precision of 
these relatively high-airmass observations, we combined our differential 
$b$ and $y$ observations into a single $(b+y)/2$ passband.  We also 
computed all differential magnitudes using the composite mean brightness 
of the three comparison stars to average out any subtle light variations 
in the comparisons.

The resulting differential magnitudes are plotted in the top panel of 
Figure~4 and are summarized in Table~5.  Even though the observing 
seasons are short, it is clear from both the table and figure that the 
mean brightness of HD~207832 varies over a range of $\sim 0.005$ mag.
In contrast, the composite mean of the three comparison stars varies over
a total range of only 0.0010 mag and has a standard deviation from the 
grand mean of only 0.00051 mag.  This demonstrates the long-term stability 
of our photometric calibrations and the reliability of the measured
variation in the yearly-mean brightness of HD~207832.  The low-level 
photometric variability in HD~207832 results from subtle changes in the 
star's magnetic activity and is in line with other solar-type stars of 
similar age \citep[see, e.g., Figure~11 of][]{Hall09}.

The presence of photospheric spots in solar-type stars allows the 
possibility of direct determination of stellar rotation periods due
to rotational modulation in the visibility of the spots and the consequent
variability in the star's brightness \citep[see, e.g.,][]{ghh2000,hfh1995}.
\citet{Queloz01} and \citet{Paulson04} have demonstrated how starspots can
result in periodic radial velocity variations that mimic the presence of 
planetary companions. We performed periodogram analyses of the individual 
four seasons of our HD~207832 photometry.  Coherent brightness variability 
with a strong rotation signal was found only in the first observing season, 
plotted in the second panel of Figure~4. We note from column 4 of 
Table~5 that the 2007 season exhibited the largest night-to-night brightness 
variability, indicating a slightly larger degree of asymmetry in the spot 
distribution and so permitted the determination of the star's rotation 
period.  The frequency spectrum of these observations in the third panel 
of Figure~4 gives a period of $17.8 \, \pm \, 0.5$ days. Much weaker signals 
of 20, 21, and 24 days were seen in 2008 and 2009 seasons, so we take 
17.8 days to be the rotation period of HD~207832.  This period is 
consistent with the star's $V \sin i=3$ km\,s$^{-1}$ and $\log {{\rm R'}_{\rm HK}}=-4.62$.  
The 2007 season data are phased with the rotation period in the bottom panel of Figure~4.  
The best-fit sine curve, also plotted in the bottom panel, has a peak-to-peak 
amplitude of 0.011 mag.

\section{Concluding Remarks}

The measurements of the radial velocities of HD~207832 suggest that two Jovian-type 
planets exist in orbit around this star. 
Given the moderate eccentricities of these two planets and the small semi-major axis of the 
inner body, it would be interesting to speculate on the origin of these objects and the possibility
of the existence of additional smaller bodies in this system.

As HD 207832 is a solar-type star with a mass only slightly smaller than that of the Sun, 
conventional wisdom holds that any Jovian-type planets associated with this star should have 
formed at large distances, beyond the protoplanetary ``ice line". This picture
suggests that planet b and (possibly) planet c were formed in regions well separated from their current orbits and 
reached their current locations either via migration, interaction with other planetary bodies, 
or a combination of both.

Given that HD~207832 is a Sun-like star, it would not be unrealistic to assume that in the past, 
this star was surrounded by a disk of planetesimals and planetary embryos. The migration of the
two planets of this system (in particular planet b) could have affected the 
dynamics of these protoplanetary bodies, scattering them out of the system and altering their accretion 
to larger (e.g., terrestrial-class) objects. Within the confines of this paradigm, it  
would be moderately unexpected to find a small planet 
orbiting in the region between planets b and c. Such an object, however,
may be able to maintain stability in the region interior to planet b.
As argued by \citet{Zhou05}, \citet{Fogg05,Fogg06,Fogg07a,Fogg07b,Fogg09}, \citet{Raymond08}, 
and many other researchers, small planets might in fact form and survive inside the orbit of a migrating 
giant body. 

To examine the possibility of the existence of terrestrial-class objects in the region between the 
two planets and interior to the orbit of planet b, we considered each planet to have an influence zone
extending from $a(1-e)-3{R_{\rm H}}$ to $a(1+e)+3{R_{\rm H}}$, 
where $R_{\rm H}$ is the planet's Hill radius. An additional object will be outside these
influence zones if its orbit is larger than 3 AU,  between 0.75 and 1.25 AU, or smaller than 0.4 AU. 
We placed a hypothetical Earth-mass planet at different distances in the two regions nearer to the star and
integrated the four-body system of the star, planets b and c, and the Earth-mass body
for 1 Myr and for different values of the hypothetical planet's semimajor axis. 
We assumed that the Earth-mass planet was initially in a circular
orbit and varied its initial semimajor axis between 0.75 AU and 1.25 AU, and between
0.05 and 0.4 AU in increments of 0.01 AU. Results indicated that, between the two planets,
in the region 0.75-1.25 AU, the orbits of all hypothetical bodies
became unstable in less than 1 Myr. The orbits of the objects between 0.3 and 0.4 AU
also became unstable in a few hundred thousand years. However, objects between 0.05 and 0.3 AU
maintained their orbits for the duration of the integration. 

Although these results indicate that a low-mass planet may be stable in a small region interior 
to the orbit of HD~207832 b,  it is not possible, based on the current observational data, to make 
a definite conclusion on the actual existence of this object. 
As mentioned before, the periodogram of the residuals of our best-fit model shows no 
significant signal, implying that the current observational data offer no significant evidence
for other planets in this system.

\acknowledgments
NH acknowledges support from the NASA Astrobiology Institute under 
Cooperative Agreement NNA09DA77 at the Institute for Astronomy, University
of Hawaii, and NASA EXOB grant NNX09AN05G. RPB acknowledges support from NASA 
OSS Grant NNX07AR40G, the NASA Keck PI program, and from the Carnegie Institution of Washington. 
GL acknowledges support from the NASA Astrobiology Institute at NASA Ames Research Center. GWH acknowledges support from 
NASA, NSF, Tennessee State University, and the State of Tennessee through its Centers of Excellence 
program. SSV gratefully acknowledges support from NSF grant AST-0307493. 
The work herein is based on observations obtained at the W. M. Keck Observatory, which is 
operated jointly by the University of California and the California Institute of Technology.
We thank the UC-Keck, NASA-Keck, and UH/IFA Time Assignment Committees for their support.   
This research has made use of the SIMBAD database, operated at CDS, Strasbourg, France. 

{\it Facilities:} Keck (HIRES).

\clearpage

\clearpage
\begin{figure}
\center
\includegraphics [scale=0.4, angle=270]{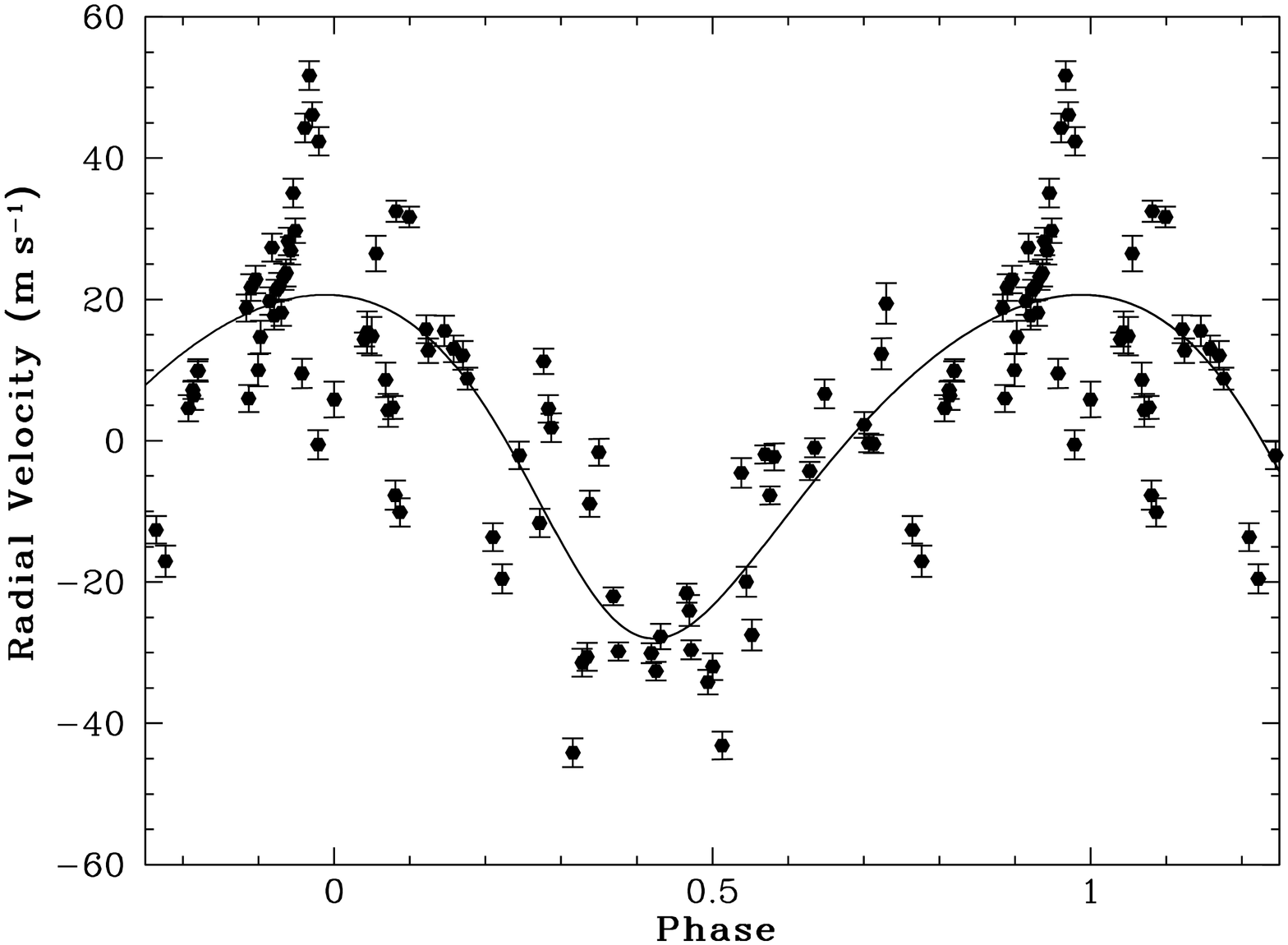}
\includegraphics [scale=0.4, angle=270]{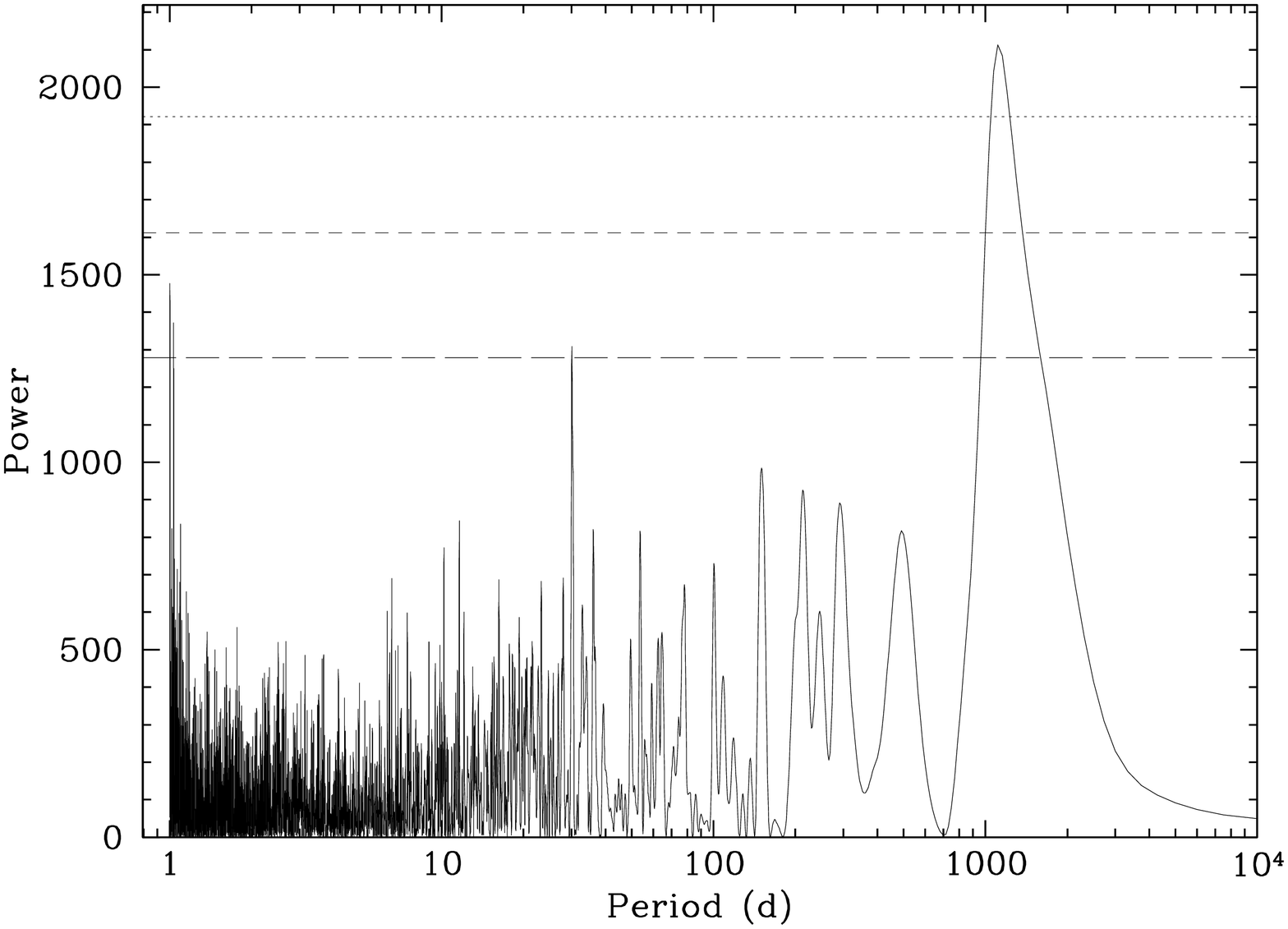}
\caption{One-planet Keplerian solution and residuals periodogram for HD~207832.
\textit{Top panel:} Phased Keplerian fit. \textit{Bottom panel:} Periodogram 
of the residuals to the one-planet best-fit solution.}\label{fig:1pfit_HD207832}
\end{figure}

\clearpage
\begin{figure}
\center
\includegraphics [scale=0.32, angle=270]{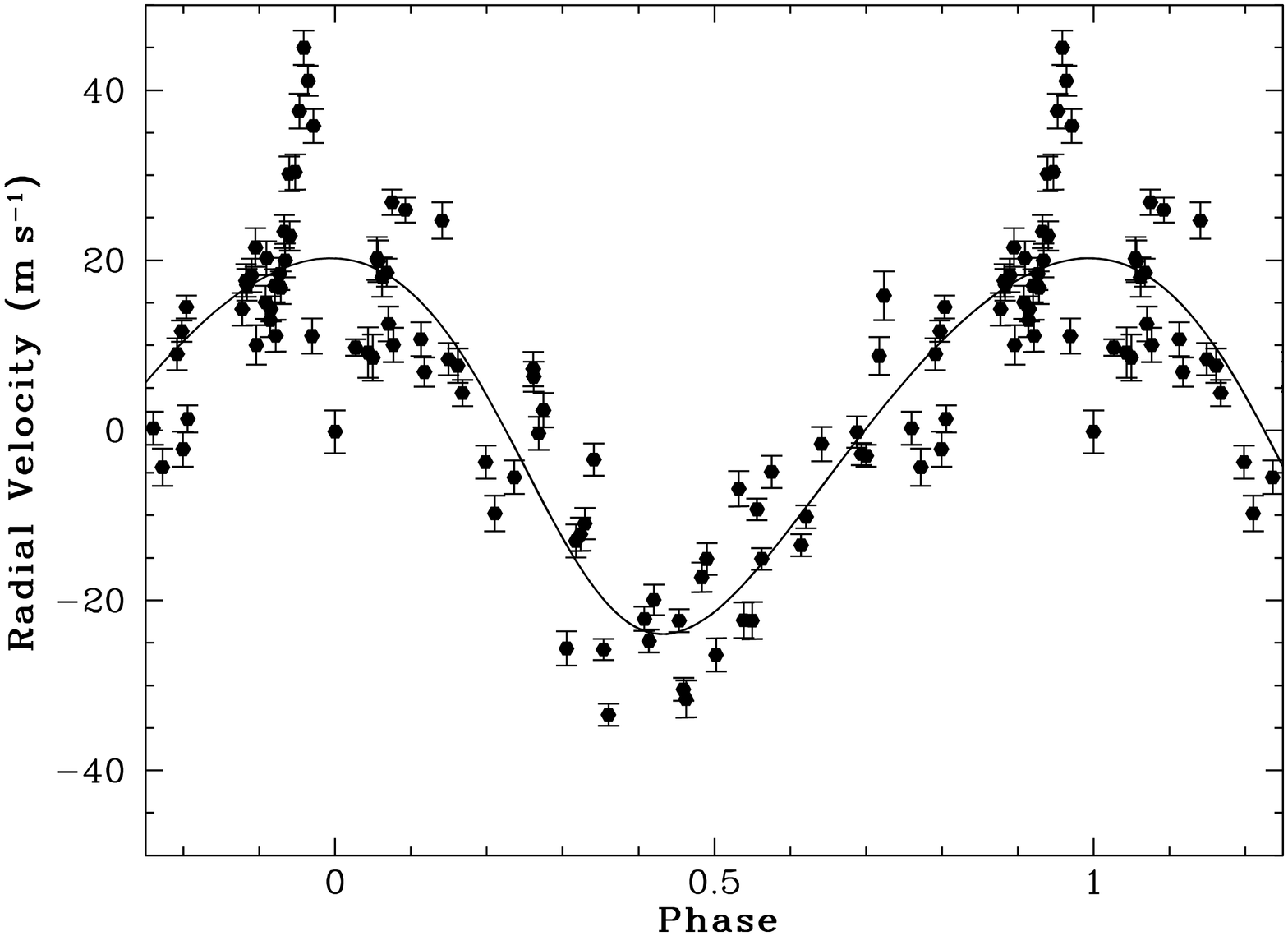}
\includegraphics [scale=0.32, angle=270]{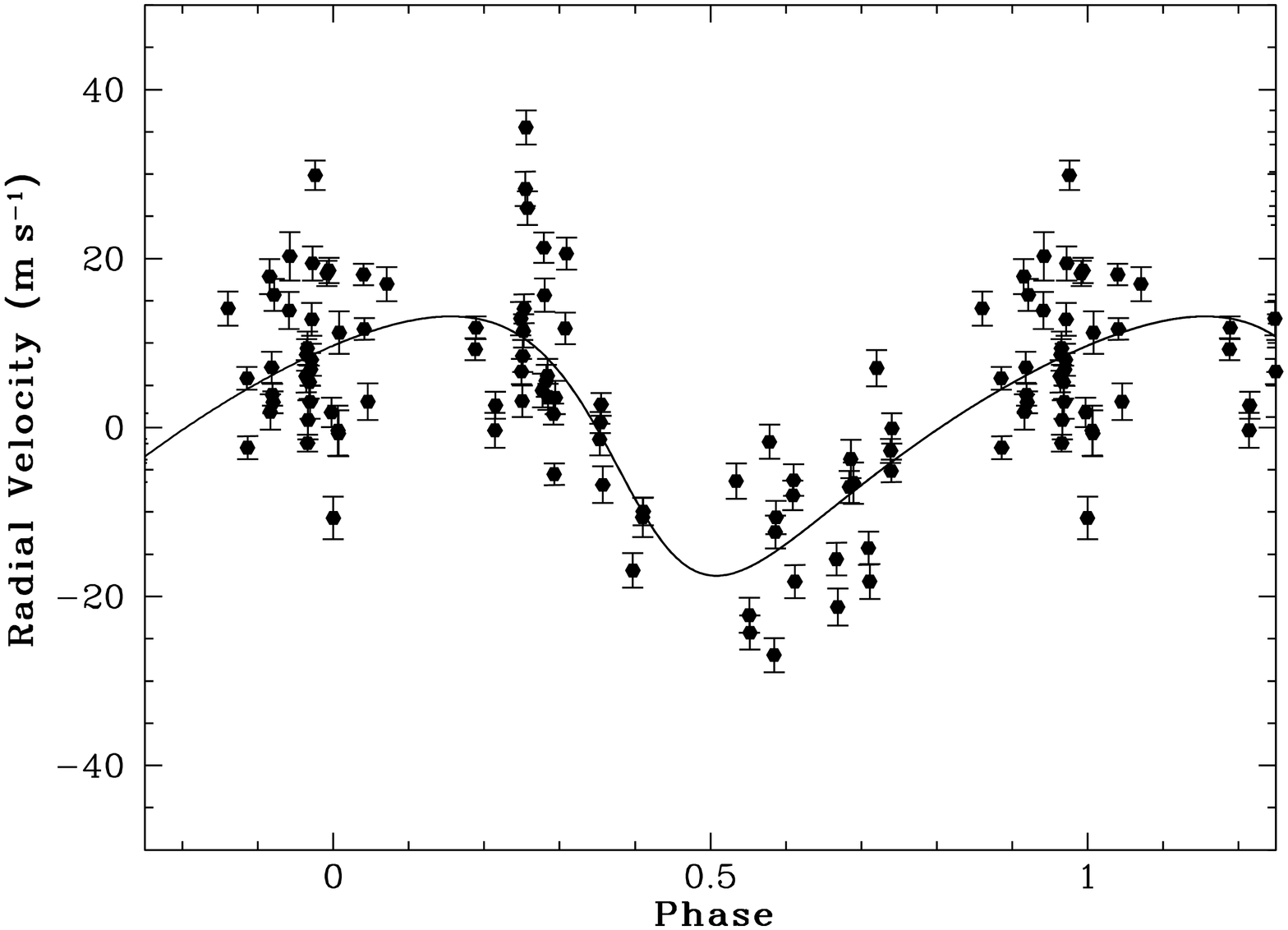}
\includegraphics [scale=0.32, angle=270]{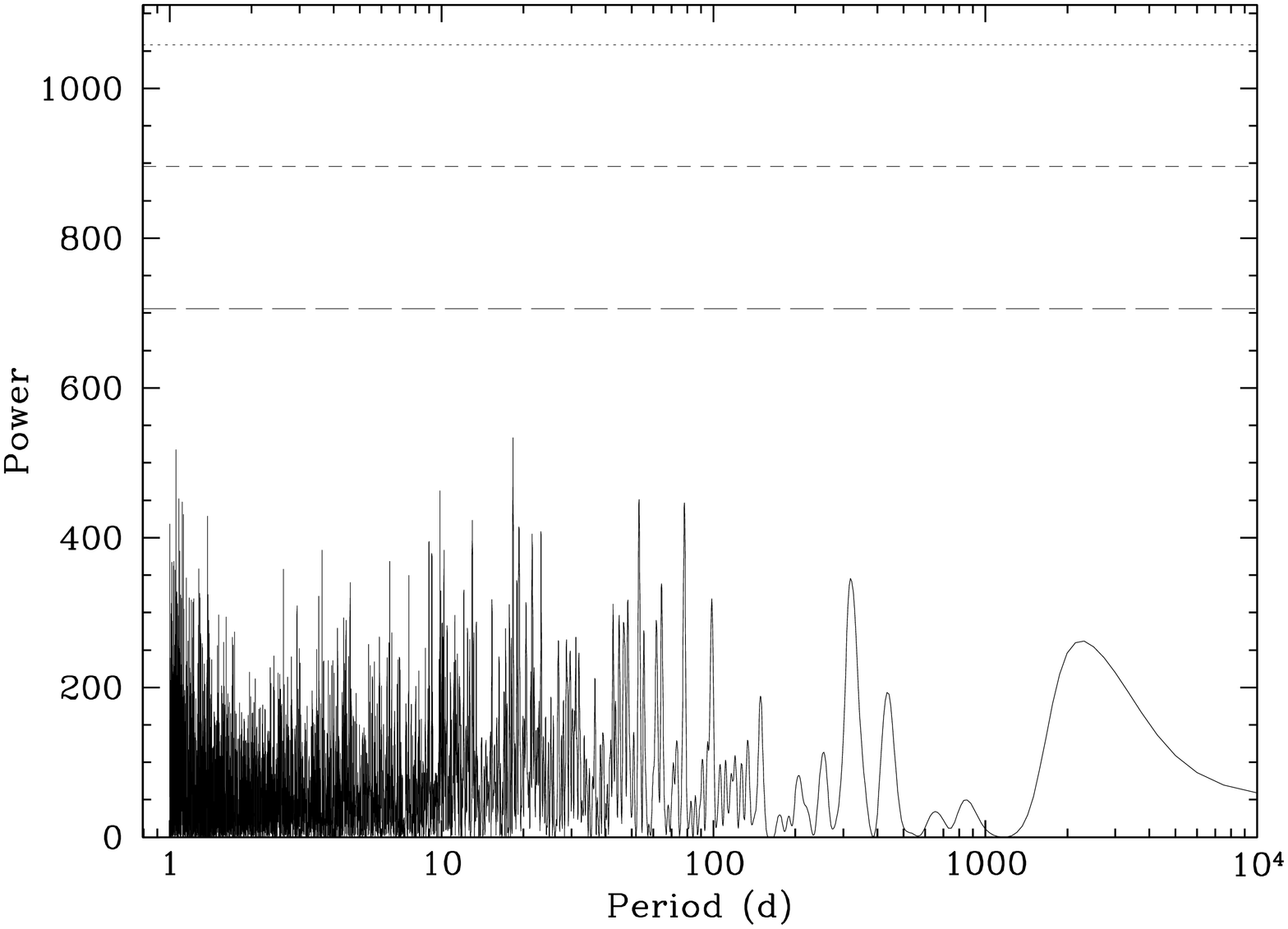}
\caption{Keplerian solutions and residuals periodogram for HD~207832.
\textit{Top panel:} Phased Keplerian fit in a system where the period of the inner planet has 
been used and the effect of the outer planet has been subtracted. \textit{Middle panel:} Phased 
Keplerian fit in a system where the period of the outer planet has 
been used and the effect of the inner planet has been subtracted.
\textit{Bottom panel:} Periodogram of the 
residuals to the two-planet best-fit solution.}\label{fig:bestfit_HD207832}
\end{figure}

\clearpage
\begin{figure}
\epsscale{0.8}
\plotone{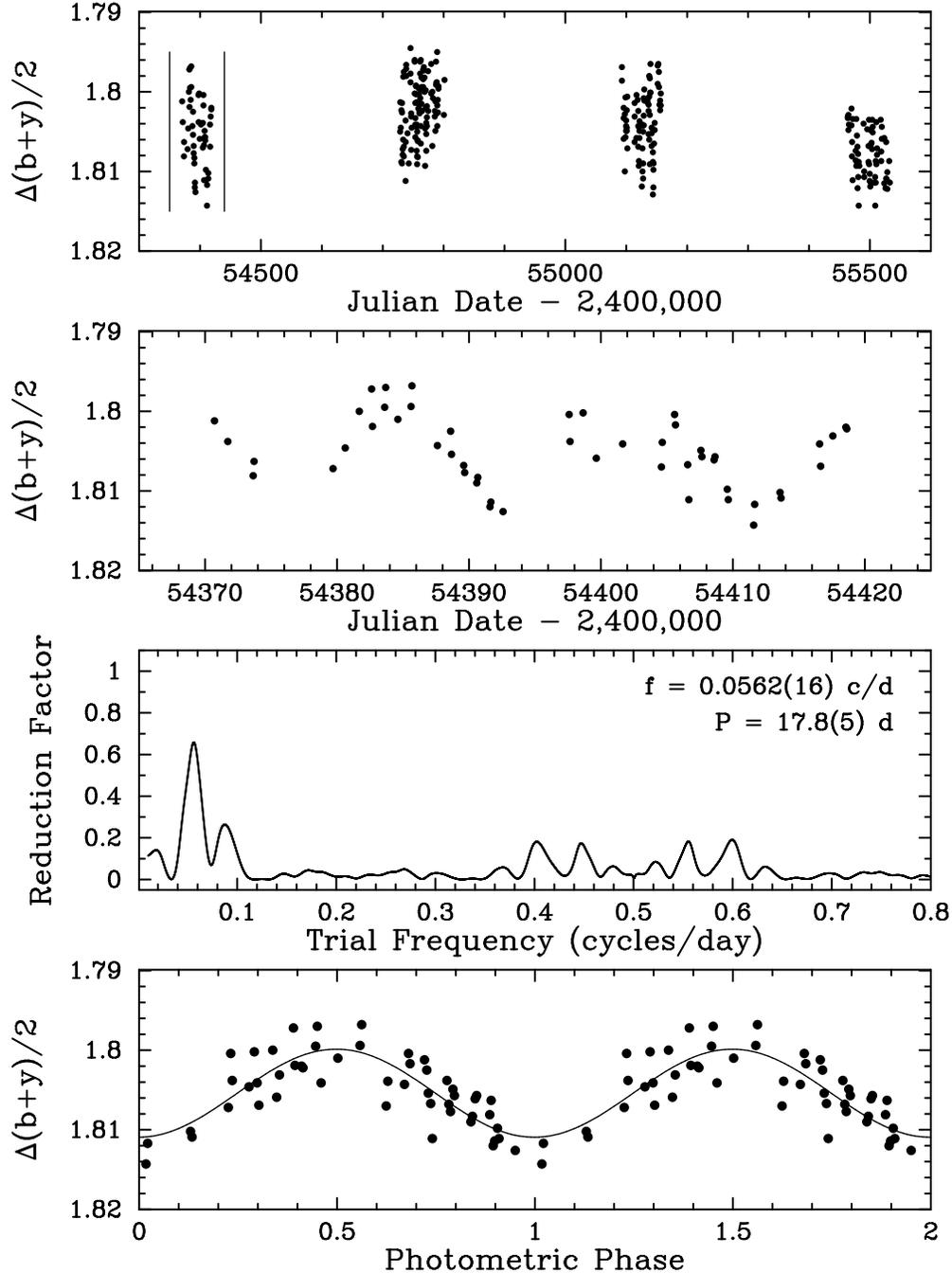}
\caption{\textit{Top}:  The 310 photometric observations of HD~207832 in the
$(b+y)/2$ passband, acquired with the T12 0.8 m APT during the 2007,
2008, 2009, and 2010 observing seasons. \textit{Second panel}:  The 2007 observing
season, set off with vertical bars in the top panel, showing the most coherent 
brightness variability due to cool starspots carried across the disk of the 
star by its rotation. \textit{Third panel}: 
Frequency spectrum of the observations in the second panel giving a best 
period of $17.8 \pm 0.5$ days. \textit{Bottom Panel}:  Plot of the data from 
panel two, phased with the star's 17.8-day rotation period. Results reveal coherent 
variability with a peak-to-peak amplitude 0.011 mag.}
\end{figure}

\clearpage

\begin{deluxetable}{lll}
\tabletypesize{\small}
\tablecaption{Stellar parameters for HD~207832}
\label{starparams}
\tablewidth{0pt}
\tablecolumns{2}
\tablehead{{Parameter}\qquad\qquad\qquad\qquad & {Value}\qquad\qquad\qquad\qquad &{Reference}}
\startdata
Spectral Type           	& G5V              & \citet{Houk82}            \\
$M_{\rm v}$                   	& 5.11$\pm$0.11    & \citet{Masana06}          \\
$B-V$                   	& 0.694	           & Hipparcos catalog         \\
$V$                     	& 8.786$\pm$0.014  & Hipparcos catalog         \\
Mass ($M_\odot$)            	& 0.94$\pm$0.10    & This work                 \\
Radius ($R_\odot$)          	& 0.901$\pm$0.056  & This work                 \\
Luminosity ($L_\odot$)   	& 0.773$\pm$0.085  & This work	               \\
Distance (pc)           	& 54.4$\pm$2.7     & \citet{VanLeeuwen07}      \\
$V \sin i$ (km\,s$^{-1}$)	& 3.0              & \citet{Nordstroem04}      \\
$S$                	& 0.258            & This work	               \\
$\log {\rm R'}_{{\rm HK}}$      & -4.62            & This work	               \\
Age (Gyr)                	& $<$ 4.5          & \citet{Holmberg09}        \\
$\rm {[Fe/H]}$                  & 0.06             & \citet{Holmberg09}        \\
$T_{{\rm eff}}$ (K)         	& 5710$\pm$81    & This work	               \\
$\log{g}$                	& 4.502$\pm$0.071  & This work	               \\
$P_{{\rm rot}}$ (days)               	& 17.8             & This work	               \\
$M_{{\rm bol},\odot}$           & 5.03$\pm$0.12    & \citet{Masana06}          \\
\enddata
\end{deluxetable}

\clearpage

\begin{deluxetable}{lcc}
\tabletypesize{\small}
\tablecaption{HIRES/Keck radial velocities for HD~207832}
\label{tab:rvdata_HD207832}
\tablewidth{0pt}
\tablehead{{JD} \qquad\qquad\qquad\qquad  & {RV (m\,s$^{-1}$)} \qquad\qquad  & {Uncertainty (m\,s$^{-1}$)}}
\startdata
2453191.07306 &    1.23 &  2.52 \\
2453198.10964 &   10.76 &  2.96 \\
2453199.10969 &   10.20 &  2.71 \\
2453200.02832 &   21.89 &  2.50 \\
2453604.05288 &  -32.09 &  2.18 \\
2453962.02939 &  -17.22 &  1.94 \\
2453964.03012 &  -21.66 &  2.20 \\
2453981.83958 &    1.37 &  1.91 \\
2453983.88414 &    5.39 &  2.27 \\
2454023.73787 &   10.94 &  2.14 \\
2454249.06943 &   -9.15 &  2.08 \\
2454250.11123 &  -24.57 &  2.11 \\
2454256.08362 &   -6.89 &  1.89 \\
2454279.05384 &    7.69 &  2.21 \\
2454280.07780 &   14.81 &  2.87 \\
2454304.98079 &   14.17 &  1.91 \\
2454305.97931 &   17.08 &  1.86 \\
2454306.98351 &   18.22 &  1.95 \\
2454308.00756 &   10.08 &  2.32 \\
2454309.97308 &   15.16 &  1.96 \\
2454310.96397 &   13.10 &  1.97 \\
2454311.96209 &   17.20 &  1.95 \\
2454312.95811 &   18.61 &  1.88 \\
2454313.95461 &   23.62 &  1.95 \\
2454314.98918 &   30.46 &  2.03 \\
2454319.05290 &   41.53 &  1.76 \\
2454336.99854 &   27.88 &  1.48 \\
2454339.83047 &   27.06 &  1.48 \\
2454343.90031 &    8.14 &  1.74 \\
2454399.76550 &  -28.65 &  2.18 \\
2454428.72442 &    2.04 &  2.02 \\
2454634.11007 &   22.74 &  1.99 \\
2454635.02452 &   16.73 &  1.48 \\
2454636.07870 &   13.52 &  1.87 \\
2454637.12316 &   19.12 &  1.91 \\
2454638.07365 &   22.29 &  1.97 \\
2454639.08741 &   25.12 &  1.72 \\
2454641.11196 &   39.66 &  2.04 \\
2454642.07952 &   47.08 &  2.03 \\
2454644.10800 &   37.76 &  2.01 \\
2454667.05175 &   11.18 &  1.99 \\
2454672.98097 &    8.41 &  1.89 \\
2454674.92028 &    7.48 &  2.03 \\
2454675.92382 &    4.18 &  1.55 \\
2454687.01449 &   -6.67 &  1.97 \\
2454702.06135 &  -13.54 &  1.85 \\
2454704.01682 &   -6.23 &  1.89 \\
2454805.78129 &   -5.17 &  2.05 \\
2454820.76637 &   -0.29 &  2.32 \\
2454821.73081 &    0.11 &  1.64 \\
2454964.12223 &    4.93 &  2.09 \\
2454984.08389 &  -12.31 &  2.06 \\
2454985.11255 &  -14.74 &  2.02 \\
2455015.01571 &  -16.25 &  2.01 \\
2455022.11297 &  -48.76 &  2.02 \\
2455024.10459 &  -36.02 &  1.95 \\
2455025.10763 &  -35.17 &  1.96 \\
2455050.98600 &  -38.79 &  1.74 \\
2455052.04273 &  -36.58 &  1.88 \\
2455054.05711 &  -47.74 &  1.95 \\
2455143.85599 &    4.02 &  2.44 \\
2455166.79983 &  -18.24 &  1.96 \\
2455168.76488 &  -24.14 &  2.09 \\
2455200.70599 &  -34.68 &  1.42 \\
2455201.70283 &  -37.22 &  1.33 \\
2455202.70257 &  -32.32 &  1.79 \\
2455341.10738 &   -2.76 &  2.01 \\
2455370.09190 &  -26.18 &  1.34 \\
2455371.02144 &  -34.21 &  1.35 \\
2455408.08145 &   -2.35 &  1.83 \\
2455409.05312 &   -4.91 &  1.29 \\
2455410.05776 &   -5.07 &  1.30 \\
2455462.94628 &    9.74 &  0.99 \\
2455548.69542 &   -6.54 &  1.27 \\
2455549.69363 &  -12.33 &  1.28 \\
2455720.05565 &   -8.89 &  1.30 \\
2455721.10910 &   -5.59 &  1.35 \\
2455750.03262 &    1.80 &  2.06 \\
2455751.03252 &    5.33 &  1.59 \\
2455824.90900 &    6.65 &  1.79 \\
2455825.96501 &   -0.09 &  1.96 \\
2455839.83583 &  -26.64 &  1.25 \\
2455840.91258 &  -34.41 &  1.29 \\
2455910.72389 &    0.00 &  1.87 \\
2455911.69779 &    2.56 &  1.24 \\
2455912.71141 &    5.28 &  1.35 \\
\enddata
\end{deluxetable}

\clearpage

\begin{deluxetable}{ll}
\tabletypesize{\small}
\tablecaption{Keplerian orbital solution for the one-planet fit}
\label{tab:fits}
\tablewidth{0pt}
\tablehead{{Parameter} & {HD~207832 b}}
\startdata
$P$ (days) & $161.82^{+0.73}_{-1.69}$    \\ 
$e$ & $0.18^{+0.15}_{-0.07}$   \\ 
$K$ (m\,s$^{-1}$) &  $24.30^{+2.96}_{-0.94}$  \\ 
$\varpi$ ($^{\circ}$) & $146.1^{+21.1}_{-115.1}$    \\
MA ($^{\circ}$) & $231.2^{+78.4}_{-28.9}$     \\
$M\sin i$ ($M_{\rm Jup}$) & 	$0.62^{+0.06}_{-0.04}$ \\ 
$a$ (AU) &  $0.569^{+0.002}_{-0.005}$  \\ 
Epoch (JD) & 	2453191.07306     \\ 
$\chi_{\nu}^2$ & \hskip 22pt  43.99  \\
RMS (m\,s$^{-1}$) & \hskip 22pt  12.33  \\
%Required Jitter (m\,s$^{-1}$)\tablenotemark{a} & \hskip 22pt 12.18  \\
\enddata
%\tablenotetext{a}{The required jitter is the value to be added in quadrature to the 
%internal uncertainties in order to make $\chi^2_{\nu}$ (nearly) equal to 1.0.}
\end{deluxetable}

\clearpage

\begin{deluxetable}{lll}
\tabletypesize{\small}
\tablecaption{Keplerian orbital solution for two-planet fit}
\label{tab:fits}
\tablewidth{0pt}
\tablehead{{Parameter} & {HD~207832 b} & {HD~207832 c}}
\startdata
$P$ (days)                & $161.97^{+0.97}_{-0.78}$ & $1155.7^{+71.9}_{-37.0}$ \\ 
$e$                       & $0.13^{+0.18}_{-0.05}$ & $0.27^{+0.22}_{-0.10}$ \\ 
$K$ (m\,s$^{-1}$)         & $22.1^{+2.7}_{-1.3}$ & $15.3^{+5.2}_{-1.0}$ \\ 
$\varpi$ ($^{\circ}$)     & $130.8^{+23.9}_{-83.4}$ & $121.6^{+32.4}_{-76.5}$ \\
MA ($^{\circ}$)           & $243.3^{+83.2}_{-30.1}$ & $211.9^{+114.4}_{-0.0}$ \\
$M\sin i$ ($M_{\rm Jup}$) & $0.56^{+0.06}_{-0.03}$ & $0.73^{+0.18}_{-0.05}$ \\ 
$a$ (AU)                  & $0.570^{+0.002}_{-0.002}$ & $2.112^{+0.087}_{-0.045}$ \\ 
Epoch (JD) & \qquad \qquad \qquad  2453191.07306 &   \\ 
$\chi_{\nu}^2$ & \qquad \qquad \qquad  \hskip 18pt  22.93 & \\
RMS (m\,s$^{-1}$) & \qquad \qquad \qquad  \hskip 20pt  8.43 & \\
%Required Jitter (m\,s$^{-1}$)\tablenotemark{a} & \qquad \qquad \qquad \hskip 20pt  8.21 & \\
\enddata
%\tablenotetext{a}{The required jitter is the value to be added in quadrature to the 
%internal uncertainties in order to make $\chi^2_{\nu}$ (nearly) equal to 1.0.}
\end{deluxetable}

\clearpage

\begin{deluxetable}{ccccc}
\tablewidth{0pt}
\tablecaption{Summary of Photometric Observations for HD~207832}
\tablehead{
\colhead{Observing} & \colhead{} & \colhead{Date Range} & 
\colhead{Sigma} & \colhead{Seasonal Mean} \\
\colhead{Season} & \colhead{$N_{\rm obs}$} & \colhead{(HJD $-$ 2,400,000)} & 
\colhead{(mag)} & \colhead{(mag)} \\
\colhead{(1)} & \colhead{(2)} & \colhead{(3)} & 
\colhead{(4)} & \colhead{(5)} 
}
\startdata
 2007 &  50 & 54370--54418 & 0.00436 & $1.80516\pm0.00062$ \\
 2008 & 111 & 54728--54801 & 0.00379 & $1.80205\pm0.00036$ \\
 2009 &  82 & 55092--55156 & 0.00386 & $1.80377\pm0.00043$ \\
 2010 &  67 & 55463--55532 & 0.00315 & $1.80778\pm0.00038$ \\
\enddata
\end{deluxetable}

\end{document}